\documentclass[12pt,a4paper]{article}
\usepackage{amsmath,amssymb,extarrows,mathtools,graphicx,subfigure,setspace}
\usepackage[all]{xy}
\usepackage{epsfig,comment,geometry,comment,color,cite,slashed}
\usepackage[active]{srcltx}
\usepackage[affil-it]{authblk}
\usepackage{fullpage}
\usepackage{subfig}
\usepackage{caption}
\numberwithin{equation}{section}
\newcommand{\be}{\begin{equation}}
\newcommand{\ee}{\end{equation}}
\newcommand{\bea}{\begin{eqnarray}}
\newcommand{\eea}{\end{eqnarray}}

\date{}
\begin{document}
	\begin{titlepage}
		{\title{\bf\fontsize{14}{15.2}{New Bi-Gravities}}}
		\vspace{.5cm}
		\author[a]{A. Akhavan}
		\author[a]{A. Naseh}
		\author[b,a]{A. Nemati}
		\author[b,a]{A. Shirzad}
		\vspace{.5cm}
		\affil[a]{ School of Particles and Accelerators, Institute for Research in Fundamental
			Sciences (IPM), 
			\hspace{.001cm} P.O.Box 19395-5531, Tehran, Iran}
		\affil[b]{ Department of Physics, Isfahan University of Technology, 
			\hspace{5.5cm} P.O.Box 84156-83111, Isfahan, Iran}
		\renewcommand\Authands{ and }
		\maketitle
		\vspace{-12cm}
		\begin{flushright}
			{\small
			}
		\end{flushright}
		\vspace{10cm}
\begin{abstract}	
We show that the problem of ghosts in critical gravity and its higher dimensional extensions can be resolved by giving dynamics to the symmetric rank two auxiliary field existing in the action of these theories. These New Bi-Gravities, at linear level around the AdS vacuum, are free of Boulware-Deser ghost, kinetic ghost and tachyonic instability within the particular range of parameters.
Moreover, we show that the energy and entropy of AdS-Schwarzschild black hole solutions of these new models are positive in the same range of parameters. This may be the sign that these new models are also free of ghost instabilities at the non-linear level.
\end{abstract}
\end{titlepage}
\setcounter{footnote}{0}
\addtocontents{toc}{\protect\setcounter{tocdepth}{4}}
\setcounter{secnumdepth}{4}
\section{Introduction}
Despite the perfect agreement of Einstein theory with observational data obtained from the Solar system, this theory is not a consistent theory for large distances. Two examples are the disability to explaining the flattening of the galaxy rotation curves \cite{Rubin:1978} and the accelerating expansion of the universe \cite{Riess:1998cb}. On the other hand, Einstein gravity  is  non-renormalizable and there is no well-known method to quantize this theory.

To resolve the first problem, so many proposals are suggested in the literature; among them, the dark matter idea seems to be more successful. In  these proposals it is common to assume the Einstein gravity is valid at all length scales but there exist invisible amounts of matter in the middle distances which reduce gravitational potential, leading to flattening rotational velocity curves. Some proposals are also presented to resolve the second problem, among them, the dark energy hypothesis is mostly accepted. The simplest form of dark energy is achieved by inserting the cosmological constant. Adopting this point of view leads to the celebrated $\Lambda$CDM (“$\Lambda$ Cold Dark Matter”) scenario which has an incredible agreement with observational data \cite{Adam:2015rua}. According to the observational data, and in high contrast with expectations from particle physics perspective, the cosmological constant is very small. In particle physics, a natural value for the vacuum energy is the mass of heaviest field in the theory, which is many order of magnitude higher than the observed cosmological constant.

One may ask why should we assume that the Einstein gravity is valid at all length scales, why the cosmological constant is so small,  and/or why we don't modify the Einstein gravity itself. To answer these questions also many proposals have appeared in the literature. Among them, in this paper we focus on  "massive gravity", i.e. a Lorentz invariant extension of Einstein gravity in which the gravity is propagated by a massive spin-2 particle. In this type of theories, the gravity becomes exponentially weak at large distances, thereby the problem of flattening of the galaxy rotation curves can be resolved. Moreover, the modified gravitational potential can lead to an accelerating expansion which its rate can be tuned by the mass term.

Fortunately, direct detection of gravitational waves in the recent experiment  GW150914, GW151226 \cite{Abbott:2016blz} by LIGO  puts an upper bound on the mass of graviton, i.e. $m_{g} < 1.2\times 10^{-22} \text{eV}$ \cite{TheLIGOScientific:2016src}\footnote{Some previous attempts to find different bounds on the mass of graviton can be found in \cite{deRham:2016nuf,Berti:2011jz} and references therein.}. The graviton mass may also link to the existence of gravitational wave polarizations.

Due to the theoretical and experimental importance of massive gravity, there is a long historical background to explore a consistent theory in this regard.
In 1939, Fierz and Pauli (FP) \cite{Fierz:1939ix} proposed a linear action for describing a free massive graviton. In the next thirty years, nothing important happened until 1970 when van Dam, Veltman \cite{vanDam:1970vg} and Zakharov \cite{Zakharov:1970cc} separately showed that the FP theory coupled to a source, in the massless limit does not reduce to the Einstein-Hilbert (EH) theory. This phenomenon is known as vDVZ discontinuity.

In 1972,  Vainshtein \cite{Vainshtein:1972sx} argued that it is not possible to find a radius, $r_{V}$ around a massive source such that the linear approximation can be trusted inside it; therefore one should consider the full non-linear theory. This opens the possibility to cure the vDVZ discontinuity by the non-linear effects. In the end of 1972, as a quick response to the Vainshtein's idea, Boulware and Deser \cite{Boulware:1973my}  argued that the non-linear massive gravities in general possess a scalar field with a wrong sign kinetic term. This unwanted mode is known as Boulware-Deser ghost.

From the point of view of effective field theory, the existence of this ghost mode is not necessarily a problem unless its mass is smaller than a UV cutoff scale. In 2002, according to this point of view, Arkani-Hamed, Georgi and Schwartz \cite{ArkaniHamed:2002sp} introduced Higgs-like mechanism to give mass to graviton. This idea was followed by Creminelli, Nicolis, Papucci and Trincherini \cite{Creminelli:2005qk}, but they showed the scalar ghost appears again in a radius much larger than the Vainshtein radius, $r_{V}$. 

In 2010, de Rham and Gabadadze \cite{deRham:2010ik} found a sign mistake in Ref. \cite{Creminelli:2005qk} and together with Tolly, proposed a consistent four-dimensional non-linear massive gravity which was free of Boulware-Deser ghost in special limits. The dRGT model was then followed and extended by Hassan and Rosen \cite{Hassan:2011vm}.  Hassan and Rosen \cite{Hassan:2011zd} then proposed a new model by giving dynamics to the auxiliary spin two field of their massive gravity model. This last model is called HR bigravity. 

On the other hand, in 2009, a parity invariant higher derivative gravitational model was suggested by Bergshoeff, Hohm and Townsend\cite{Bergshoeff:2009hq} in three-dimensional space-time which at the linear level contained a FP massive spin-2 mode, a massless one and no Boulware-Deser ghost. This theory which is called New Massive Gravity (NMG) is described by
\bea\label{NewMassiveGravity}
I_{\text{NMG}} = \frac{1}{16\pi G_{3}}\int d^{3}x
\sqrt{-g}\left(R-2\lambda -\frac{1}{m^{2}}(R^{\mu\nu}R_{\mu\nu}-\frac{3}{8}R^{2})\right).
\eea 
Unfortunately, the sign of kinetic terms of the massless and massive spin-2 particles are always opposite in this theory. This kind of instability is called  the kinetic ghost.  Moreover, the dual central charges have the same sign as the kinetic term of the massless spin-2 particle. This apparent problem is called "bulk-boundary clash". This problem was finally solved in 2013 \cite{Bergshoeff:2013xma}  by introducing an extension of NMG in vierbein formalism. Very recently another extension of NMG  in metric formalism \cite{Akhavan:2016hju} is presented, which has the same benefits as NMG and is free of the problem of bulk-boundary clash  .

Definitely, the NMG model provides an alternative way through a new consistent massive gravity in four-dimensional space-time along with dRGT and HR bigravity models. In this direction, L$\ddot{\text{u}}$ and Pope in 2011 proposed \cite{Lu:2011zki}  the four-dimensional version of NMG called critical gravity as
\bea\label{CriticalGravity}
I_{\text{CG}} = \frac{1}{16\pi G}\int d^{4}x \sqrt{-g}\left(R
-2\Lambda -\frac{1}{2m^{2}}(R^{\mu\nu}R_{\mu\nu}-\frac{1}{3}R^2)\right),
\eea
where $R_{\mu\nu}$, $R$, $\Lambda$ and $m^{2}$ are Ricci tensor, Ricci scalar, cosmological constant and a mass parameter, respectively. This theory and its higher dimensional extensions \cite{Bergshoeff:2011ri}, also suffer from the same kinetic ghost problem as in the NMG model. The aim of this paper is to explore a possibility of resolving  this problem for these theories in the manner that the Boulware-Deser mode remains non-dynamic. In this way, we introduce a new four dimensional (and higher dimensional) massive gravity model that differs from the well-known dRGT and HR bigravity models\footnote{Recently some new bi-gravity models \cite{Li:2015izu} are proposed which the Boulware-Deser ghost is absent in them. But these models suffer from kinetic ghosts.}. 

In the next section, we introduce our model and explain its differences with dRGT and HR bigravity models. In section.\ref{DB}, we study the AdS-wave solutions of this model. In section.\ref{DC}, we study this model at the linear level and show that it is free of ghosts and tachyons. In sections.\ref{DD}, we provide an evidence for consistency of this model at the full non-linear level. This evidence is positivity of energy and entropy of AdS-Schwarzschild black hole solution in the same parameters range where the model is ghost free and tachyon free. In the last section, we discuss about some additional calculations  to check more the consistency of this model. Apart the above historical review, since the motivation for this work also comes from the AdS/CFT correspondence, we do our calculations by considering AdS space. However, to complete our discussion, we explain some results regarding the flat space in appendix \ref{Details of Linearization} as well as details of linearization for arbitrary background in D dimensions. We will show that our model is also consistent around possible flat solutions.
\section{The Model}\label{DA}
In this paper, the main idea to resolve the kinetic ghost problem of critical gravity, (\ref{CriticalGravity}), comes from the recent paper \cite{Akhavan:2016hju} in which a new consistent three-dimensional massive gravity model is proposed. To see this idea in D-dimensional spacetime, let us start with the D-dimensional analogous of NMG which is, in fact, the critical gravity action. This action described in terms of auxiliary symmetric field $f_{\mu \nu}$ reads  \cite{Bergshoeff:2011ri} 
\bea\label{CGD}
I = \frac{1}{16\pi G}\hspace{-1mm}\int d^{D}x \sqrt{-g}\left(
R[g]-2\Lambda_{g} +\frac{1}{D-2} f_{\mu\nu}\mathcal{G}^{\mu\nu}[g]+\frac{m^2}{4 (D-2)}(\tilde{f}^{\mu\nu}f_{\mu\nu}-\tilde{f}^{2})\right)\hspace{-.9mm},
\eea
where $\mathcal{G}_{\mu\nu}$ is the Einstein tensor (due to the metric $g_{\mu\nu}$) and $m^{2}$
is a mass parameter. Here we use a notation in which $\tilde{f}^{\mu\nu} \equiv g^{\mu\alpha}g^{\nu\beta}f_{\alpha\beta}$, $\tilde{f} \equiv g^{\mu\nu} f_{\mu\nu}$. Solving the equations of motion for the field $f_{\mu\nu}$ gives
\bea
f_{\mu\nu} = -\frac{2}{m^2}\Big(R_{\mu\nu}[g]-\frac{1}{2 (D-1)} R[g] g_{\mu\nu}\Big).
\eea
Substituting back this expression in (\ref{CGD}) gives the action of critical gravity \cite{Bergshoeff:2011ri, Deser:2011xc}, in its higher derivative form,
\bea\label{critical gravityD}
I = \frac{1}{16\pi G}\hspace{-1mm}\int d^{D}x \sqrt{-g}\left(R
-2\Lambda_{g} -\frac{1}{m^2 (D-2)} (R^{\mu\nu}R_{\mu\nu}- \frac{D}{4(D-1)} R^2\hspace{1mm})\right)\hspace{-.9mm}.
\eea
One can find the dynamical degrees of freedom, at the linearized level, by performing the linearization of the theory (\ref{CGD}) around a maximally
symmetric vacuum with $AdS_{D}$ geometry.

For generic values of the parameters, it is shown that the theory (\ref{CGD}), around this background, describes one
massless spin-2 and  one massive spin-2 particle with mass \cite{Bergshoeff:2011ri}
\bea
M^{2} = (D-2)\left(m^{2}+\frac{\Lambda}{(D-1)}\right).
\eea
Moreover, the kinetic terms of massless and massive modes have opposite signs and therefore there is no way to get rid of ghosts. To avoid these ghosts, one can choose a special value for $\Lambda$ \cite{Bergshoeff:2011ri} as
\bea
m^{2}+\frac{\Lambda}{(D-1)} =0.
\eea
Now another problem arises. The linearized equation of motion for this value of $\Lambda$ reads
\bea
\left(\Box -\frac{4\Lambda}{(D-1)(D-2)}\right)^{2}h_{\mu\nu} =0,
\eea
which its quadratic nature implies logarithmic modes. Hence, the dual field theory is a LCFT which is non-unitary\cite{Alishahiha:2011yb,Bergshoeff:2011ri,Johansson:2012fs}. To conclude, the theory (\ref{CGD}) always contains ghosts.

Based on the idea of Ref.\cite{Akhavan:2016hju}, to find an extension for the theory (\ref{CGD}) which has a consistent unitary and tachyon free spectrum, we promote the auxiliary field $f_{\mu\nu}$ to a dynamical field by adding kinetic and cosmological terms for it. Fortunately, the Boulware-Deser ghost remains non-dynamical in this way. We present the new action as
\bea\label{bnmgD}
&& I = \frac{1}{16\pi G}\int d^{D}x \sqrt{-g}\left(
R[g]-2\Lambda_{g} +\frac{1}{D-2} f_{\mu\nu}\mathcal{G}^{\mu\nu}[g]+\frac{m^2}{4 (D-2)}(\tilde{f}^{\mu\nu}f_{\mu\nu}-\tilde{f}^{2})\right)+\cr\nonumber\\
&&\hspace{3cm}+\frac{1}{16\pi \tilde{G}}
\int d^{D}x\sqrt{-f}\hspace{.5mm}\bigg(R[f]-2\Lambda_{f}\bigg),
\eea
where $\tilde{G}$ and $\Lambda_{f}$ are the Newton constant and cosmological constant for the field $f_{\mu\nu}$, respectively. In the subsequent sections, we study different aspects of this model and show that, at the linearized level around AdS vacuum, it is free from ghosts and tachyonic instabilities. Before that, let us emphasize on two differences between this model and the well-known massive gravity models, i.e. dRGT model and HR bigravity. First, these models are based on the idea of extending the FP mass term through a potential which does not contain derivative terms. However, we will show that  in the theory (\ref{bnmgD}), the graviton mass has also contributions from the derivative term $f_{\mu\nu}\mathcal{G}^{\mu\nu}[g]$.

The second and most important difference is explained by rewriting the HR bigravity model, in its higher derivative form. As is discussed in \cite{Hassan:2013pca} for this model, one can determine $f_{\mu\nu}$ algebraically in term of $g_{\mu\nu}$ and its curvatures $R_{\mu\nu}[g]$. In general, the solution $f_{\mu\nu}(g)$ is a perturbative expansion in powers of $\frac{1}{m^{2}}R_{\mu\nu}[g]$, where $m^{2}$ sets the scale of the FP mass. Using this perturbative solution to eliminate $f_{\mu\nu}$ from the HR bigravity action, one can obtain the higher derivative gravity action,  $I_{\text{HD}}[g]=I_{\text{HR-BiG}}[g,f(g)]$, which at the four-derivative level reads \cite{Hassan:2013pca}
\bea\label{Truncated}
I_{\text{HD}}[g]= \frac{1}{16\pi G}\int d^{4}x \sqrt{-g}\left(R
-2\tilde{\Lambda} -\frac{1}{2\tilde{m}^2}(R^{\mu\nu}R_{\mu\nu}-\frac{1}{3 }R^2)\right)+\mathcal{O}(\frac{1}{m^{4}}).
\eea
Neglecting higher order terms, $\mathcal{O}(\frac{1}{m^{4}})$, this action is the critical gravity action (\ref{CriticalGravity}). This point is exactly where the difference between the model HR bigravity and our model (\ref{bnmgD}) is clarified. Hassan, Schmidt-May and von Strauss \cite{Hassan:2013pca} have shown that the action (\ref{Truncated}), without $\mathcal{O}(\frac{1}{m^{4}})$ terms, gives a massive spin-2 particle which is ghost and its mass differs from the value in associated HR bigravity model. They argued that the appearance of the ghost with a different mass is the artifact of truncating the original higher derivative theory to a four-derivative action (\ref{Truncated}); therefore to resolve this discrepancy all the higher order terms $\mathcal{O}(\frac{1}{m^{n}});n\geq 4$ should be added. In the current work, instead of adding all those higher derivative terms, we rewrite the action (\ref{Truncated}) in its D-dimensional auxiliary form (\ref{CGD}) and remove the ghost by promoting the auxiliary field in the model to a dynamical field. We show that this way also gives a consistent massive gravity model.

Let's begin with the equations of motion for the two fields $g_{\mu\nu}$ and $f_{\mu\nu}$ which can be obtained by varying the action (\ref{bnmgD}) as  
\bea\label{e.o.m.gD}
&& \mathcal{G}[g]_{\mu\nu}+\Lambda_{g}g_{\mu\nu}=\frac{1}{(D-2)}\left(T_{\mu\nu}^{(1)}[g]+T_{\mu\nu}^{(2)}[g]\hspace{.5mm}\right),\cr\nonumber\\
&&\mathcal{G}_{\mu\nu}[f]+\Lambda_{f} f_{\mu\nu}=\frac{1}{(D-2)}T_{\mu\nu}[f],
\eea
where $\kappa = \frac{G}{\tilde{G}}$ and
\bea\label{e.o.m.gD.1}
&& T_{\mu\nu}^{(1)}[g] =-\frac{m^2}{2}
\bigg[\tilde{f}^{\rho}_{\mu}f_{\nu\rho}-\tilde{f} f_{\mu\nu}-\frac{1}{4}g_{\mu\nu}
(\tilde{f}^{\rho\sigma}f_{\rho\sigma}-\tilde{f}^{2})\bigg] ,\cr \nonumber\\
&&  T_{\mu\nu}^{(2)}[g] =\bigg[-2\tilde{f}_{(\mu}
\hspace{.25mm}^{\rho}\mathcal{G}[g]_{\nu)\rho}
-\frac{1}{2}f_{\mu\nu}R[g]+\frac{1}{2}\tilde{f} R_{\mu\nu}[g]+\frac{1}{2}g_{\mu\nu}
f_{\rho\sigma}\mathcal{G}[g]^{\rho\sigma}-\cr \nonumber\\
&& -\frac{1}{2}\bigg(\nabla^{2}[g]f_{\mu\nu}-2\nabla[g]^{\rho}\nabla[g]_{(\mu}f_{\nu)\rho}+\nabla[g]_{\mu}\nabla[g]_{\nu}\tilde{f} +(\nabla[g]^{\rho}\nabla[g]^{\sigma}
f_{\rho\sigma}-\nabla^{2}[g]\tilde{f})g_{\mu\nu}\bigg)\bigg],\cr \nonumber\\
&&  T_{\mu\nu}[f] =\frac{1}{\kappa }\sqrt{\frac{g}{f}}\hspace{1mm}\bigg[f_{\alpha\mu}f_{\beta\nu}
\mathcal{G}[g]^{\alpha\beta}+\frac{m^2}{2}\left(g^{\sigma\alpha}
g^{\tau\beta}-g^{\sigma\tau}g^{\alpha\beta}\right)\left
(f_{\sigma\tau}f_{\alpha\mu}f_{\beta\nu}\right)
\bigg].\nonumber\\
\eea
Note that we did not call the $g_{\mu\nu}$ or $f_{\mu\nu}$ as the metric. Because for the moment it is not clear which of them is the source for the energy-momentum tensor or equivalently which of them corresponds to the massless graviton. In fact, as we see in section.\ref{DC}, the real metric may be a proper combination of both at linear level.
\section{AdS Wave Solutions}\label{DB}
In this section, we present one type of solutions, known as "AdS wave" solutions (since they are a special kind of gravitational waves propagating along the AdS spacetime). In general they can be written as
\bea\label{AdSwave}
g_{\mu\nu}= g_{\mu\nu}^{\text{AdS}} -F\hspace{.5mm}k_{\mu}k_{\nu},
\eea
where $k_{\mu}$ is a null geodesic field, and $F$ is a function which depends on the dynamics of the gravitational theory. Albeit they look like perturbative excitations around the AdS spacetime, one should note that they are  solutions of the full non-linear equations of motion. This statement can be understood by the null characteristic of the vector field $k_{\mu}$.

The  AdS wave solutions are studied as a preliminary test of unitarity of the underlying theory. As we will see, the form of the function $F$ is closely related to the particle content of a theory. Non-unitarity of a theory may be showed up in the AdS wave solutions. However, if this test is passed, there is no guarantee for the theory to be unitary and one still needs more consistency checks.

In this section, we present the general AdS wave solutions for Eqs.(\ref{e.o.m.gD}). These are important since they solve  the linearized equations of motion, as well. Due to the complexity of Eqs. (\ref{e.o.m.gD}), one should be careful about the form of the ansatz given.  
We consider the following ansatz for the fields $g_{\mu\nu} $ and $f_{\mu\nu}$
\bea
&& ds^{2}_{g}=\frac{\ell_{g}^{2}}{r^{2}}\Big(dr^{2}+dx_{i}^{2}-2 dx^{+}dx^{-}-G(r,x^{+}) dx^{+ 2}\Big),\cr \nonumber\\
&& ds^{2}_{f}=\frac{\ell_{f}^{2}}{r^{2}}\Big(dr^{2}+dx_{i}^{2}-2 dx^{+}dx^{-}-F(r,x^{+}) dx^{+ 2}\Big).
\eea
Plugging this ansatz in the equation of motion of $ g_{\mu\nu} $ gives the following relations
\bea\label{eom1D} 
\Lambda_{g} \ell_{g}^2+\frac{1}{2}(D-1)(D-2)-\frac{1}{4} \frac{\ell_{f}^{2}}{\ell_{g}^{2}} (D-1)(D-4) \Big[1-\frac{m^2}{2 (D-2)}\ell_{f}^{2}\Big]=0,
\eea
and
\bea\label{de1D} 
&&\Big((D-6)\frac{\ell_{f}^{2}}{2\ell_{g}^{2}} -(D-2)\Big)\left[\frac{\partial^{2}G}
{\partial r^{2}}-(D-2)\frac{1}{r}\frac{\partial G}
{\partial r}\right]+\frac{\ell_{f}^{2}}{\ell_{g}^{2}}\left[\frac
{\partial^{2} F}{\partial r^{2}}
-(D-2)\frac{1}{r}\frac{\partial F}{\partial r}
\right]-\cr \nonumber\\
&&\hspace{3.25cm}-\frac{\ell_{f}^{2}}{\ell_{g}^{2}}(D-2) (D-1-m^2 \ell_{f}^{2}) \left[\frac{G}{r^{2}}-\frac{ F}{r^{2}}\right]=0.
\eea
Similarly the equation of motion of $ f_{\mu\nu} $ gives 
\bea\label{eom2D}
\Lambda_{f} \ell_{f}^{2}+\frac{1}{2}(D-1)(D-2)+\frac{1}{2\kappa}\frac{(D-1)}{ (D-2)}\Big(2-D+m^2 \ell_{f}^{2}\Big)\Big(\frac{\ell_{g}}{\ell_{f}}\Big)^{D-4}=0,
\eea
and
\bea\label{de2D} 
&&\left[\frac{\partial^{2}G}
{\partial r^{2}}-(D-2)\frac{1}{r}\frac{\partial G}
{\partial r}\right]- \kappa (D-2) \Big(\frac{\ell_{f}}{\ell_{g}}\Big)^{D-4} \left[\frac{\partial^{2} F}{\partial r^{2}}-(D-2)\frac{1}{r}\frac{\partial F}{\partial r}
\right]+\cr\nonumber\\
&&\hspace{2.3cm}+(D-2) \Big( D-1-m^2\ell_{f}^{2} \Big)\left[\frac{G}{r^{2}}-\frac{ F}{r^{2}}\right]=0.
\eea
A useful class of AdS wave solutions for the equations (\ref{eom1D})-(\ref{de2D}) is given by proportionality condition $ \ell_{f}^{2}=\gamma \ell_{g}^{2}\equiv \gamma \ell^{2}$. That is because, as we will show in section.\ref{DC}, the model (\ref{bnmgD}) has a well-defined mass spectrum around two proportional AdS$_{D}$, where the fluctuations $\delta g_{\mu\nu}$, $\delta f_{\mu\nu}$ decompose into a massless spin-2 mode and a FP massive spin-2 mode. By using the proportionality condition and assuming power law dependence of the functions $F$ and $G$, with respect to the radial coordinate "$r$", the most general solution of the Eqs. (\ref{de1D}) and (\ref{de2D}) reads
\bea\label{G1} 
&& G(r,x^+)=\mathrm{g}_{1}(x^+)+\mathrm{g}_{2}(x^+)\hspace{.5mm} r^{D-1}+\mathrm{g}_{3}(x^+)\hspace{.5mm}r^{\frac{D-1}{2}(1+\sqrt{1+A}\hspace{.5mm})}
+\mathrm{g}_{4}(x^+)\hspace{.5mm}r^{\frac{D-1}{2}(1-\sqrt{1+A}\hspace{.5mm})},\cr \nonumber\\
&& F(r,x^+)=\mathrm{g}_{1}(x^+)+\mathrm{g}_{2}(x^+)\hspace{.5mm}r^{D-1}+\beta\hspace{.5mm} \mathrm{g}_{3}(x^+)\hspace{.5mm}r^{\frac{D-1}{2}(1+\sqrt{1+A}\hspace{.5mm})}
+\beta\hspace{.5mm} \mathrm{g}_{4}(x^+)\hspace{.5mm}r^{\frac{D-1}{2}(1-\sqrt{1+A}\hspace{.5mm})},\cr \nonumber\\
\eea
where $\mathrm{g}_{i}(x^+) $'s are arbitrary functions of $x^+$ and
 \bea\label{A-ads wave}
&&A=-4\frac{(D-2)^2 \big(D-1-m^2\ell_{g}^{2} \gamma\big)\Big(\gamma-2-2\kappa\gamma ^{D/2-1}\Big)}{(D-1)^2\Big(2\gamma +\kappa \gamma ^{\frac{D}{2}-2}(D-2)\big(-6 \gamma +(\gamma-2) D+4\big)\Big)},\nonumber\\
&&\beta=-\frac{1}{2}\frac{\big(-4 \gamma +(\gamma -2) D+4\big)}{\big(\gamma-(D-2) \kappa  \gamma ^{D/2-1}\big)}.
\eea
To explore the particle content of the theory (\ref{bnmgD}) by using the AdS wave solutions (\ref{G1}), note that the AdS wave solutions are also solutions of linearized equations of motion which are closely related to the particle content of a theory. In the following, we explore this subject for $ D=4 $. Comparing to (\ref{G1}), the AdS waves of Einstein-Hilbert theory just contain functions $f_{1}$ and $f_{2}$. On the other hand, the Einstein-Hilbert theory contains only the massless spin-2 particle. Hence, the AdS wave solutions (\ref{G1}) mean that besides the massless spin-2 particle, the theory (\ref{bnmgD}) has other particles in its spectrum, related to the functions $f_{3}$ and $f_{4}$. To get information about these new modes a good, although naive, way is comparing the AdS wave solutions (\ref{G1}) with solutions of the wave equation for a massive spin-2 particle, i.e. 
\bea\label{waveequation}
(\Box +\frac{2}{\ell^{2}}-\mathbb{M}^{2})h_{\mu\nu} =0,
\eea
where the D'Alembert operator is defined with an AdS$_{4}$ background with radius $\ell$. In general, Eq.(\ref{waveequation}) has two independent solutions which can be combined as follows
\bea\label{test}
h_{\mu\nu} \sim \frac{\ell^{2}}{r^{2}}\left[a(x^{i})\hspace{1mm}  r^{\frac{3}{2}\big(1-\sqrt{1+\frac{4}{9}\mathbb{M}^{2}\ell^{2}}\hspace{.5mm}\big)}+b(x^{i})\hspace{1mm} r^{\frac{3}{2}\big(1+\sqrt{1+\frac{4}{9}\mathbb{M}^{2}\ell^{2}}\hspace{.5mm}\big)}\hspace{.5mm}\right],
\eea 
where $a,b$ are arbitrary functions of spatial coordinates. It is clear that for the massless particle, $\mathbb{M}^{2}=0$, we have $h_{\mu\nu} \sim \frac{\ell^{2}}{r^{2}}\left[a(x^{i}) +b(x^{i})\hspace{.5mm}r^{3}\right]$. Therefore the AdS wave solutions (\ref{G1}), in comparison with (\ref{test}), tells us that the theory (\ref{bnmgD}) at least has  one massless spin-2 and one massive spin-2 particle with mass
\bea\label{m2A}
\mathbb{M}^{2} = \frac{9}{4\ell^{2}}A.
\eea  
We emphasize on the word ``at least", because the wave equation of massive scalar with mass $\mathbb{M}^{2}$ on the AdS$_{4}$ spacetime has the same solution as (\ref{test}). However, by this naive analysis, we are not sure that the theory (\ref{bnmgD}) contains such a scalar particle. To assure about it, one needs to find the action of quadratic fluctuations, which is the subject of section.\ref{DC}. 

Now the unitarity, which here means the non-tachyonic nature of excitations, impose the condition $\mathbb{M}^{2} \geq -\frac{9}{4\ell^{2}}$, where the lower bound is known as Breiteinlohner-Freedman (BF) bound in AdS \cite{Breitenlohner:1982bm}. Hence, for any value $A \geq -1$, the theory (\ref{bnmgD}) is free of tachyonic spin-2 particles. This last condition can constrain the parameters of the theory according to (\ref{A-ads wave}). However, note that by this analysis we can not say whether or not the theory contains the kinetic ghosts.

Let's get back to the solutions (\ref{G1}). It may happen that for some special values of parameters, we have logarithmic AdS-wave solutions; which some of them must be avoided because of unitarity.  The log-solutions arise at $A=0$ and $A=-1$. The case $ A=0 $ implies
\bea\label{dangerous}
\gamma=\frac{1}{m^2\ell_{g}^{2}}(D-1),\hspace{.5cm}\text{or}\hspace{.5cm}1-\frac{\gamma}{2}+\kappa \gamma ^{\frac{D}{2}-1}=0,
\eea
which converts the massive spin-2 mode to massless one. This phenomenon which is common in all critical gravities \cite{Alishahiha:2011yb},\cite{Lu:2011zki}, \cite{Grumiller:2013at} shows  that the theory for the values (\ref{dangerous}) may be non-unitary. The reason is that, for these values the AdS wave solutions of theory (\ref{bnmgD}) should be written as
\bea\label{Log0D} 
\hspace{-1.5cm}ds^{2}_{g} = \frac{\ell_{g}^{2}}{r^{2}}\bigg(dr^{2}+dx_i^{2}-2 dx^+ dx^{-}-\mathbf{g}(r,x^{+}) dx^{+ 2}\bigg),\cr\nonumber\\
\hspace{-1.5cm}ds^{2}_{f} = \frac{\gamma \ell_{g}^{2}}{r^{2}} 
\bigg(dr^{2}+dx_i^{2}-2 dx^+ dx^- -\mathbf{f}(r,x^{+}) dx^{+ 2}\bigg),
\eea
with
\bea
&& \mathbf{g}(r,x^{+}) = \tilde{G}_{0}[x^+]\log(r)
+G_{0}[x^+]+\tilde{G}_{D-1}[x^+]r^{D-1}\log(r)
+G_{D-1}[x^+]r^{D-1}, \cr \nonumber\\
&& \mathbf{f}(r,x^{+}) =\tilde{F}_{0}[x^+]\log(r)
+F_{0}[x^+]+\tilde{F}_{D-1}[x^+]r^{D-1}\log(r)
+F _{D-1}[x^+]r^{D-1}.\nonumber
\eea
The presence of leading $\text{Log}$ term means that the dual theory is a logarithmic conformal field theory (LCFT), i.e. a non-unitary theory. This is also understandable from the fact that the theory (\ref{bnmgD}) for the values (\ref{dangerous}) has two massless spin-2 particles in its spectrum.
For the case $A=-1$, the solutions are of the form (\ref{Log0D}) with
\bea
&& \mathbf{g}(r,x^{+}) = G_{0}[x^+]+\tilde{G}_{\frac{D-1}{2}}[x^+]r^{\frac{D-1}{2}}\log(r)+G_{\frac{D-1}{2}}[x^+]r^{\frac{D-1}{2}}
+G_{D-1}[x^+]r^{D-1}, \cr \nonumber\\
&& \mathbf{f}(r,x^{+}) =F_{0}[x^+]+\tilde{F}_{\frac{D-1}{2}}[x^+] r^{\frac{D-1}{2}}\log(r)
+F _{\frac{D-1}{2}}[x^+]r^{\frac{D-1}{2}}+F_{D-1}[x^+]r^{D-1}.\nonumber
\eea
However, the appearance of these $\text{Log}$ solutions is not the sign that the dual theory is a LCFT since the  mass $\mathbb{M}^{2}$ is non-zero. Therefore the dangerous values of the parameters which causes non-unitary logarithmic solutions, and should be avoided, are given by Eq.(\ref{dangerous}).
\section{Linearization}\label{DC}
In section.\ref{DB}, we showed that the theory (\ref{bnmgD}) has vacuum solution with two proportional AdS metrics. In the following, we determine the spectrum of propagating modes of the theory (\ref{bnmgD}) around this vacuum solution, where $ \bar{f}_{\mu\nu} =\gamma \bar{g}_{\mu\nu} $. Inserting these  AdS solutions in the equations of motion (\ref{e.o.m.gD}) gives (similar to Eqs.(\ref{eom1D}) and (\ref{eom2D}))
\bea\label{BeomD} 
&& \Lambda_{g} \ell^2+\frac{1}{2}(D-1)(D-2)-\frac{\gamma}{4}(D-1)(D-4) \Big[1-\frac{m^2}{2 (D-2)}\gamma\ell^{2}\Big]=0,\cr \nonumber\\
&& \Lambda_{f}\gamma\ell^{2}+\frac{1}{2}(D-1)(D-2)+\frac{1}{2\kappa}\frac{(D-1)}{(D-2)}\Big(2-D+\gamma m^2 \ell^{2}\Big)\gamma^{(2-\frac{D}{2})}=0,
\eea
where  $\ell$ is the radius of $\bar{g}_{\mu\nu}$. Consider the general form of fluctuations as follows
\bea
g_{\mu\nu}=\bar{g}_{\mu\nu}+h_{\mu\nu},~~~f_{\mu\nu}=\gamma(\bar{g}_{\mu\nu}+\rho_{\mu\nu}).
\eea
We expand various tensor terms of the action (\ref{bnmgD}) up to second order in the perturbations $ h_{\mu\nu} $ and $ \rho_{\mu\nu} $. These expansions are given in appendix \ref{Details of Linearization}.
By this way, the quadratic action in perturbations emerges as 
\bea\label{linearD}
&& S^{(2)}[h_{\mu\nu}, \rho_{\mu\nu}]=\frac{1}{16\pi G}\hspace{-1mm}\int\hspace{-1mm} d^{D}x\sqrt{-\bar{g}}\hspace{1mm}\Bigg\{\frac{2(D-2)-(D-6)\gamma}{2(D- 2)}  h^{\mu\nu}(\mathbb{G} h)_{\mu\nu}\hspace{-1mm}+\hspace{-1mm}\kappa \gamma^{\frac{D}{2}-1} \rho^{\mu\nu}(\mathbb{G} \rho)_{\mu\nu}\cr\nonumber\\
&&\hspace{5cm}-\frac{2 \gamma}{(D-2)}h^{\mu\nu}(\mathbb{G}\rho)_{\mu\nu}+(h-\rho).(h-\rho)
\Bigg\},
\eea
where $\mathbb{G}$ is the Pauli-Fierz operator on the curved $ AdS_{D} $ background which is defined as 
\bea\label{PFD}
&& p^{\mu\nu}(\mathbb{G} q)_{\mu\nu}\equiv -\dfrac{1}{4} p_{\nu\rho ;\mu} q^{\nu\rho ;\mu}+\dfrac{1}{2} p_{\mu\nu ;\rho} q^{\rho\nu ;\mu}-\dfrac{1}{4} p_{;\mu} {q^{\mu\nu}}_{;\nu}-\dfrac{1}{4} q_{;\mu} {p^{\mu\nu}}_{;\nu}
+\frac{1}{4}p_{;\mu}q^{;\mu}-\cr \nonumber\\
&&\hspace{2.3cm}-\frac{(D-1)}{2 \ell^{2}}(p^{\mu\nu}q_{\mu\nu}-\frac{1}{2}pq),
\eea
and the dot product is defined as
\bea\label{dot}
A\cdot A\equiv \chi\; A_{\mu\nu}A^{\mu\nu}-\xi\; A^2 ,
\eea
with 
\bea\label{dot product}
\chi=\frac{\gamma}{4\ell^2}(D-1-\gamma m^2\ell^2),~~~~~~\xi=\frac{\gamma}{8\ell^2}\frac{\Big(D^2+2-(D-3)\gamma m^{2}\ell^2-3 D\Big)}{(D-2)}.
\eea
To investigate the presence of ghost instabilities, we need to omit the cross terms in the quadratic action (\ref{linearD}). For this reason, it is useful to utilize  
a new basis for the fluctuations as follows
\be\label{basis}
h= a_1h^{(0)}+a_2\hspace{.5mm}h^{(m)},\hspace{1cm}\rho= \hspace{.5mm}h^{(0)}+ \hspace{.5mm}h^{(m)}.
\ee
Hence, the quadratic action reads
\bea\label{S1D}
&&\hspace{-1cm}S^{(2)}[h^{(0)}_{\mu\nu},h^{(m)}_{\mu\nu}]=\dfrac{1}{16\pi G}\int d^{D}x\sqrt{-\bar{g}}\hspace{1mm}\Bigg\{A h^{(0)\mu\nu}(\mathbb{G} h^{(0)})_{\mu\nu}+B h^{(m)\mu\nu}(\mathbb{G} h^{(m)})_{\mu\nu}+C h^{(m)\mu\nu}(\mathbb{G}h^{(0)})_{\mu\nu}\cr\nonumber\\
&&\hspace{2cm}+\chi (a_1-1)^2\left( h^{(0)}_{\mu\nu} h^{(0) \mu\nu}-\frac{\xi}{\chi} h^{(0)2}\right)+\chi (a_2-1)^2 \left( h^{(m)}_{\mu\nu} h^{(m) \mu\nu}-\frac{\xi}{\chi} h^{(m)2}\right)\cr \nonumber\\
&&\hspace{2cm}+2\chi (a_1-1) (a_2-1)\left( h^{(0)}_{\mu\nu} h^{(m) \mu\nu} -\frac{\xi}{\chi}h^{(0)} h^{(m)}\right)
\Bigg\},
\eea
where the precise expressions for $A,B,C$ are as below
\bea\label{AD}
&& A=\kappa \gamma^{\frac{D}{2}-1}+\frac{a_1}{2(D-2)}\Bigl(2a_1(D-2)-
\bigl[4+a_1(D-6)\bigr] \gamma\Bigr) ,\hspace{.5cm}B=A\hspace{.5mm}[a_1\rightarrow a_2],\cr \nonumber\\
&& C=2\kappa \gamma^{\frac{D}{2}-1}+\frac{1}{(D-2)}\bigg(2a_1 a_2 (D-2)-\bigl[2(a_1+a_2)+a_1a_2(D-6)\bigr]\gamma\bigg),
\eea
The absence of scalar ghost (Boulware-Deser ghost \cite{Boulware:1973my}) in propagating modes $h^{(0)}$ and $h^{(m)}$ implies that $\xi= \chi=1$. The last condition gives
\bea\label{ghost free aprameter}
\gamma= \frac{1}{m^2\ell^2}(D-2).
\eea
Using this value for $\gamma$, gives 
\bea\label{M3,4D}
&& C=2\kappa\hspace{1mm}(\frac{D-2}{m^2\ell^2})^{\frac{D}{2}-1}-\frac{1}{m^2\ell^2}\bigg(2 (a_1+a_2)+a_1 a_2 (D-6-2m^2\ell^2)\bigg).
\eea
To have decoupled modes in relation (\ref{S1D}), we should imply $ C=0 $ in the first line and the third line should vanish, by that we get 
\bea\label{a1a2}
a_1=1\hspace{.5cm}\text{or}\hspace{.5cm}a_2=1\hspace{.5cm}\text{or}\hspace{.5cm}a_1=a_2=1.
\eea
We discard the case $a_1=a_2=1$ since for this case vanishing of $C$ implies that 
$$2m^2 \ell^2-(D-2)+2\kappa  m^2\ell^2 (\frac{D-2}{m^2\ell^2})^{\frac{D}{2}-1} =0, $$ 
which is exactly the condition for the appearance of logarithmic solution (\ref{Log0D}). Moreover, the symmetry of the model under interchange of $a_1$ and $a_2$
implies that the two cases $ a_1=1$ and $a_2=1$ are physically equivalent. Therefore, in the following we restrict ourselves to the case $a_1=1$. Vanishing of the coefficient C for this case implies 
\bea\label{a_2D}
a_2=-\frac{2}{(D-2)}\frac{\Big(D-2-\kappa m^4\ell ^4 (\frac{D-2}{m^2\ell^2})^{\frac{D}{2}} \
	\Big)}{(D-4-2m^2\ell^2)}.
\eea
To this end, the proper basis in which the theory is scalar ghost free and we have two decoupled propagating modes is the following
\bea
h=h^{(0)}-\frac{2}{(D-2)}\frac{\Big(D-2-\kappa m^4\ell ^4 (\frac{D-2}{m^2\ell^2})^{\frac{D}{2}} \
	\Big)}{(D-4-2m^2\ell^2)}\hspace{.5mm}h^{(m)},\hspace{.5cm}\rho =h^{(0)}+h^{(m)}.
\eea
In this basis, the action (\ref{S1D}) becomes
\bea\label{QuadraticActionD}
&&\hspace{-1cm}S^{(2)}[h^{(0)},h^{(m)}]=\frac{1}{16\pi G}\int d^{D}x \sqrt{-\bar{g}}\hspace{1mm}
\bigg[\mathbb{A}_{0}\hspace{.5mm}h^{(0)\mu\nu}
(\mathbb{G}h^{(0)})_{\mu\nu}\hspace{1mm}+\cr \nonumber\\
&&\hspace{2.3cm}+ \mathbb{A}_{m}\left\{h^{(m)\mu\nu}
(\mathbb{G}h^{(m)})_{\mu\nu}-\frac{\mathbb{M}^{2}}{4}\bigg( h^{(m)\mu\nu}h^{(m)}_{\mu\nu}-(h^{(m)})^{2}\bigg)\right\} \bigg],
\eea
where
\bea\label{AAD}
&&\mathbb{A}_{0}=1-\frac{1}{2m^2\ell^2}(D-2) + \kappa\hspace{.5mm}(\frac{D-2}{m^2\ell^2})^{\frac{D}{2}-1},\cr\nonumber\\
&&\mathbb{A}_{m}=-2\hspace{.5mm}
\frac{\Big(2+\kappa m^2\ell^2(D-6-2m^2\ell^2)
	(\frac{D-2}{m^2\ell^2})^{\frac{D}{2}-1}\Big)}{
	(D-4-2m^2\ell^2)^2}\mathbb{A}_{0}
,\cr\nonumber\\[2.5mm]
&&\mathbb{M}^{2}=-\frac{4m^{2}(D-2)}{(D-4-2m^{2}\ell^{2})^{2}}\frac{\mathbb{A}_{0}^{2}}{\mathbb{A}_{m}}.
\eea
It is illustrative to take a closer look at the above results in the limits, $\kappa \rightarrow 0$ and $m^{2}\ell^{2}\rightarrow 1$, where the theory (\ref{bnmgD}) reduces to the critical gravity (\ref{critical gravityD}). In these limits, in $D=4$ the quadratic action (\ref{QuadraticActionD}) vanishes which seems to be in contradiction with the critical gravity in the linearized level. However, the correct way to taking the limits, $\kappa \rightarrow 0$ and $m^{2}\ell^{2}\rightarrow 1$, is imposing them in Eq.(\ref{M3,4D}) noticing (\ref{a1a2}) to find $a_{1}=a_{2}=1$ upon demanding the absence of the mixing terms. As stated before, this is discarded due to appearance of $\text{Log}$ solutions (\ref{Log0D}). 

\begin{figure}[h!]
	\begin{center}
		\centering
		\captionsetup{justification=centering,margin=3cm}
		\vspace{1cm}
	    \includegraphics[height=6cm, width=6cm]{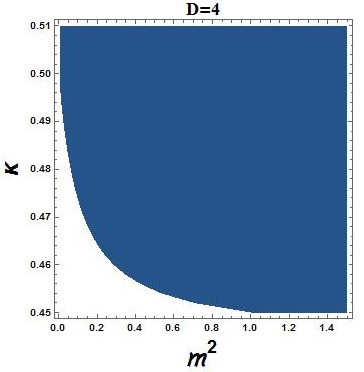}
		\includegraphics[height=6cm, width=6cm]{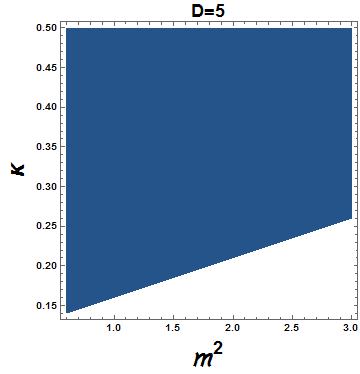}
		\includegraphics[height=6cm, width=6cm]{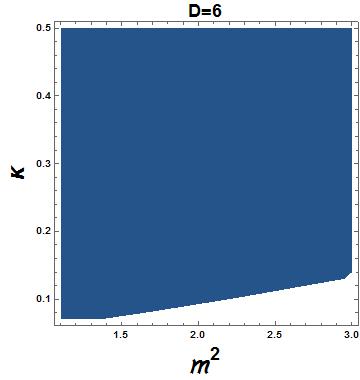}
		\includegraphics[height=6cm, width=6cm]{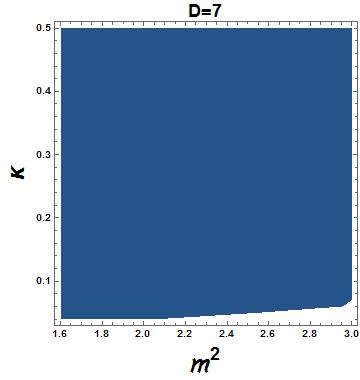}
		\caption{Unitary region for $m^{2}$ and $\kappa$. Note that here we set $\ell=1 $.}\label{fig1}
	\end{center}
\end{figure}

\begin{figure}[h!]
	\begin{center}
		\vspace{1cm}
		\centering
		\captionsetup{justification=centering,margin=1.5cm}
	    \includegraphics[height=6cm, width=7cm]{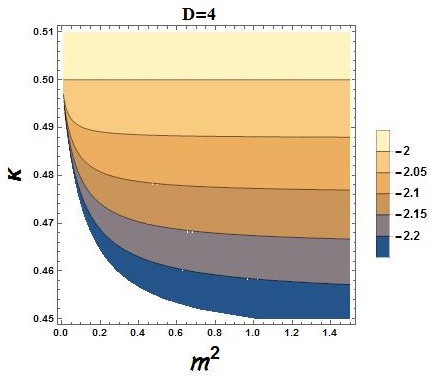}
		\includegraphics[height=6cm, width=7cm]{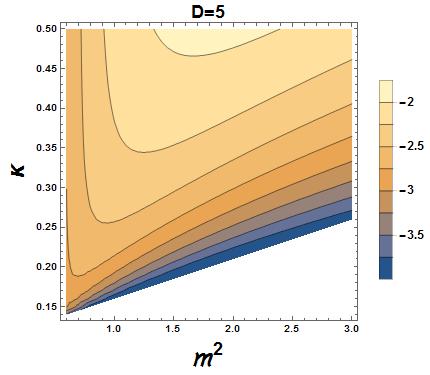}
		\includegraphics[height=6cm, width=7cm]{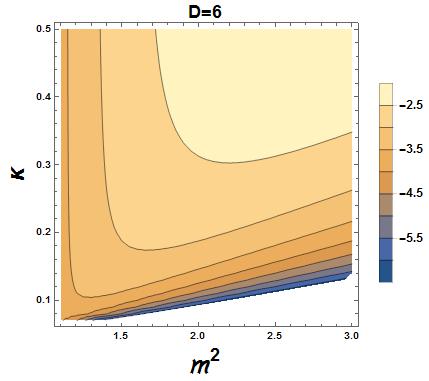}
		\includegraphics[height=6cm, width=7cm]{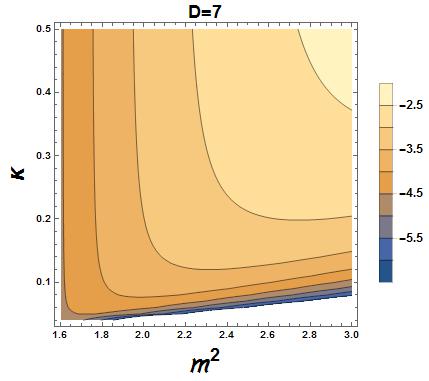}
		\caption{Values of $\mathbb{M}^{2}$ within the unitary region for $m^{2}$ and $\kappa$. Note that here we set $ \ell=1 $.}\label{fig2}
	\end{center}
\end{figure}

Ghost free condition together with absence of tachyonic mode implies that
\bea
\mathbb{A}_{0}>0,\hspace{.5cm}\mathbb{A}_{m}>0,\hspace{.5cm} \mathbb{M}^{2}\geq -\frac{(D-1)^2}{4 \ell^2}.
\eea
According to the above conditions, for $D=4$, $ 5 $, $6$ and $7$, we presented the allowed values of $(m^{2},\kappa)$ in Fig.\ref{fig1}. For these values, the massive mode acquires different masses in the range $-\frac{(D-1)^2}{4 \ell^2} \leq{M}^{2} <0$ as depicted in Fig.\ref{fig2}.

However, in the allowed region, Fig.\ref{fig1}, it exists a subregion for $D=4$ in which the mass of the spin-2 particle becomes a special value, $\mathbb{M}^{2}=-\frac{2}{\ell^{2}}$. The theory (\ref{bnmgD}) at this subregion has an interesting feature that is the appearance of a new gauge symmetry which eliminates
the helicity zero part of the massive spin-2 field, leaving behind only four propagating modes. These four remaining degrees of freedom collectively present a Partially Massless spin-2 particle (PM mode)\cite{Deser:1983mm} and \cite{Deser:1983tm}. That new gauge symmetry is
\bea
\delta_{\xi}h_{\mu\nu}^{(m)} = \left(\bar{\nabla}_{\mu}\bar{\nabla}_{\nu}-\frac{1}{\ell^{2}}\bar{g}_{\mu\nu}\right)\xi(x),
\eea
where the $\bar{g}_{\mu\nu}$ is AdS$_{4}$ background and the covariant derivative is defined also by this metric. Hence, the quadratic action (\ref{QuadraticActionD}) in that subregion reduces to
\bea\label{PMaction}
&&S^{(2)}_{\text{PM}}[h^{(0)},h^{(m)}] =\frac{1}{16\pi G} \int d^{4}x \sqrt{-\bar{g}}\hspace{1mm}
\bigg[h^{(0) \mu\nu}
(\mathbb{G}h^{(0)})_{\mu\nu}\hspace{1mm}+\cr \nonumber\\
&&\hspace{2cm}+\frac{1}{m^{2}\ell^{2}}\hspace{.75mm}\left\{ h^{(m) \mu\nu}
(\mathbb{G}h^{(m)})_{\mu\nu}+\frac{1}{2\ell^2}\bigg( h^{(m)\mu\nu}
h^{(m)}_{\mu\nu}-(h^{(m)})^{2}\bigg)\right\} \bigg].
\eea
The point should be noticed is that this action is the same as four-dimensional Conformal Gravity\cite{Lu:2011ks} at the linearized level. This means that the Weyl symmetry, in that special subregion, comes back to the theory (\ref{bnmgD}) for $ D=4 $, at least at the linearized level. If this structure can be extended to all orders, then the theory (\ref{bnmgD}) can provide a non-linear theory for PM particles. 

Before closing this section, let us try to answer a natural question which might be asked: How much the analysis in this section depends to the background metric (vacuum solution). The crucial point that should be considered to answer this question is explicitly the absence of Riemann tensor in the action (\ref{bnmgD}) and equations of motion (\ref{e.o.m.gD}). This means that around the backgrounds (vacuum solutions) where they have a same Ricci tensor and Ricci-Scalar tensor, the spectrum of particles and unitary regions would be exactly the same. An example of such backgrounds are the AdS$_{D}$ spacetime and Schwarzschild-AdS black hole where the Ricci tensor and Ricci-Scalar tensor for both of them are $-\frac{D-1}{\ell^{2}}g_{\mu\nu}$ and $-\frac{D (D-1)}{\ell^{2}}$, respectively\footnote{We have checked the linearized analysis for these two different backgrounds explicitly and found the same spectrum and unitary regions.}. In next section, we check the existence of these black hole solutions for the model (\ref{bnmgD}).
\section{Schwarzschild-AdS Black Hole Solutions}\label{DD}
In this section, we present one of the black hole solutions for the theory (\ref{bnmgD}). We consider two following proportional  Schwarzschild-AdS solutions 
\bea\label{AdSSchD}
ds^2_g=-F(r) dt^2+\frac{1}{F(r)}dr^2+r^2 d\Omega^{2}_{D-2},\hspace{.8cm}ds^2_f=\gamma ds^2_g,
\eea
where $F(r)= 1+\frac{r^2}{\ell^2}-(\frac{\mu}{r})^{D-3}$. The parameter $\mu$ is related to the mass of the black hole and can be expressed in terms of the horizon radius $r_h$ as follows
\bea\label{AdSSchD2}
\mu=r_h\Big(1+\frac{r_h^2}{\ell^2}\Big)^{\frac{1}{D-3}}.
\eea
The metric (\ref{AdSSchD}) reduces to D-dimensional anti-de Sitter spacetimes with radius of curvature $\ell$ for $\mu=0$, and converts to the standard Schwarzschild solution for $\ell \rightarrow \infty$. Substituting this ansatz  (\ref{AdSSchD}) into the equations of motion (\ref{e.o.m.gD}) and (\ref{e.o.m.gD.1}) gives the same equations as (\ref{BeomD}) for two proportional AdS. The reason is that the Ricci tensor and Ricci-Scalar tensor, which are the only tensors present in the equations of motion (\ref{e.o.m.gD}) and (\ref{e.o.m.gD.1}), are the same for  Schwarzschild-AdS black hole and pure AdS spacetime.  
\subsection{Energy and Entropy of Black Hole Solutions}\label{DE} 
In this subsection, we find the energy and entropy of black hole solutions (\ref{AdSSchD}) using the ``renormalized on-shell action". To have a well-defined variational principle, depending on the boundary conditions, one needs the appropriate Gibbons-Hawking terms. The black hole solutions (\ref{AdSSchD}) are obtained according to the Dirichlet boundary condition. However, the theory (\ref{bnmgD}) may have other solutions which are obtained by different boundary conditions, such as $\text{Log}$ solutions (\ref{Log0D}) where the variational principle, as well as the Gibbons-Hawking terms, should be modified appropriately.   Here, we are interested in some ranges of parameters of the theory (\ref{bnmgD}) where these solutions do not exist. On the other hand, the on-shell action, in general, may contain divergences which should be removed by adding suitable counterterms. This is what we mean by  the word "renormalized".

The standard way to do this is using the free energy function. Choosing the Dirichlet boundary condition for the Schwarzschild-AdS black hole solutions (\ref{AdSSchD}), the corresponding boundary terms which may violate the variational principle emerge from the following terms 
\bea
\delta I_{\delta\partial_{r}g,\delta\partial_{r}f} = \frac{1}{16\pi G}
\int d^{D}x\left(\sqrt{-g} 
\delta R[g]+ \frac{1}{D-2} \sqrt{-g}f_{\mu\nu}
\delta \mathcal{G}^{\mu\nu}[g]+\kappa \sqrt{-f} \delta R[f]\right),
\eea
which can be written as
\be\label{DeltaD}
\delta I_{\delta\partial_{r}g,\delta\partial_{r}f} = \frac{1}{16\pi G}
\int d^{D}x\left(\sqrt{-g}\mathcal
{A}_{\mu\nu}\delta R^{\mu\nu}[g]+
\kappa \sqrt{-f}
f_{\mu\nu}\delta R^{\mu\nu}[f]\right),
\ee
where
\bea
\mathcal{A}_{\mu\nu}=g_{\mu\nu}+\frac{1}{D-2} (f_{\mu\nu}-\frac{1}{2}
f_{\alpha\beta}g^{\alpha\beta}g_{\mu\nu}).
\eea
In general, the appropriate Gibbons-Hawking term for the variations (\ref{DeltaD}) is difficult to find. We can simplify the problem by considering  the particular subspace of the  space of solutions in which the two fields are proportional, i.e. $f_{\mu\nu}=\gamma g_{\mu\nu}$. For this case, the variation terms (\ref{DeltaD}) can be simplified as 
\be
\delta I_{\delta\partial_{r}g,\delta\partial_{r}f} 
= \frac{1}{16\pi G}\int d^{D}x\left(\sqrt{-g}(1-\frac{\gamma}{2} ) g_{\mu\nu}\delta 
R^{\mu\nu}[g]+\kappa \sqrt{-f}
f_{\mu\nu}\delta R^{\mu\nu}[f]\right),
\ee
from which the proper Gibbons-Hawking terms can be suggested as follows 
\bea\label{GHD}
I_{\text{GH}} =- \frac{2(1-\frac{ \gamma}{2})}
{16\pi G}\int d^{D-1}x\sqrt{-\eta_g}\;
K[g]-\frac{2 }{16\pi \tilde{G}}\int 
d^{D-1}x \sqrt{-\eta_f} K[f],
\eea
where ${\eta_g}_{ij}$ and ${\eta_f}_{ij}$ are the induced metrics, on the boundary, associated 
with the metrics $g$ and $f$ and $K[g]=\eta_g^{ij}K[g]_{ij},
K[f]=\eta_{f}^{ij}K[f]_{ij}$. To find the general form of Gibbons-Hawking term one can use the methodology of Ref. \cite{Hohm:2010jc}. 

To this end, a well-defined variational principle implies that  the  Gibbons-Hawking terms of Eq.(\ref{GHD}) should be added to the original bulk action \eqref{bnmgD} 
(named $I_0$). Let us substitute the Schwarzschild-AdS black hole solutions (\ref{AdSSchD}) in this total action. By changing to the Euclidean signature, $ t\rightarrow i\tau$, and integrating over $r (r_{h}\rightarrow \cal R) $, $\tau (0\rightarrow \beta)$ and angular parameters we arrive at 
\bea
I_{0}+I_{\text{GH}}=\frac{a}{4 G} \frac{(D-2) (\sqrt{\pi})^{D-3}}{\Gamma (\frac{D-1}{2})}\left(\frac{{\cal R}^{D-1}}{\ell^2}+{\cal R}^{D-3} +b\right),
\eea
where ${\cal R}\gg r_h$ is a cutoff and
\bea
a=\beta (1-\frac{\gamma}{2}+\kappa \gamma^{\frac{D}{2}-1}),\hspace{.75cm}
b=-\frac{r_h^{D-3}}{2 (D-2)}\bigg (D-1+ (D-3) \frac{r_h^2}{\ell^2}\bigg).
\eea
It is clear that the above on-shell action is divergent due to the infinite volume limit, so appropriate counterterms are needed to remove the divergent terms. One can easily check that the suitable counterterms are
\bea\label{IctD}
I_{\text{ct}}=\frac{1}{16 \pi G}\int d^{D-1}x\sqrt{\eta _g}\hspace{1mm}\bigg( c_{1}+ c_{2} R[\eta _g]+c_3 R^{ij}[\eta _g] R_{ij}[\eta _g]+c_4 R[\eta _g]^2+...\bigg),
\eea
with
\bea\label{IctD.1}
&& c_{1}=-\frac{2 (D-2)}{ \ell}\big(1-\frac{\gamma}{2} +\kappa \gamma ^{\frac{D}{2}-1}\big),\hspace{.4cm}c_{2} =\frac{\ell^{2}}{2(D-2)(D-3)}\hspace{.1mm}c_{1},\cr \nonumber\\
&&c_{3}=-\bigg((D-2)+\frac{(D-5)}{2(D-1)} \ell c_{1}\bigg)c_{4},\hspace{.3cm} c_4=\frac{(D-1)\ell^3}{4(D-2)(D-3)^2(D-5)}.\nonumber\\
\eea
It should be noticed that the counterterms in (\ref{IctD}) render the renormalized action finite up to $D=7$. For $D=4$ and $ 5 $
one needs just $c_{1}$ and $c_{2}$ terms and for $D=6$ and $7$ one needs  all of $c_{i}$ ($i < 5$) terms. As a check point the counterterm action (\ref{IctD.1}) for the case $\gamma =0$ becomes
\bea
&& I_{\text{ct}}=-\frac{1}{16 \pi G}\int d^{D-1}x\sqrt{\eta _g}\hspace{1mm}\bigg[ \frac{2}{\ell}(D-2)+ \frac{\ell}{(D-3)}R[\eta _g]\hspace{.5mm}+\cr \nonumber\\ 
&&\hspace{2cm} +\frac{\ell^{3}}{(D-3)^{2}(D-5)}\bigg(R^{ij}[\eta _g]R_{ij}[\eta _g]-\frac{(D-1)}{4(D-2)}R^{2}[\eta _g]\bigg)+...\bigg],
\eea
which is the standard counterterms for Einstein-Hilbert action \cite{deHaro:2000vlm, Balasubramanian:1999re} by identifying $D=d+1$\footnote{Note that our curvature convention differs by a minus sign with the  convention of \cite{deHaro:2000vlm}.}. Putting everything together, the renormalized on-shell action reads
\bea
&& I_{\text{ren}} = I_{0}+I_{\text{GH}}+I_{\text{ct}}=\frac{\tilde{a}}{16 \pi G}\hspace{.5mm}  \big(1-\frac{\gamma}{2}+\kappa\hspace{.1mm} \gamma^{\frac{D}{2}-1}\big),\cr \nonumber\\
&&\hspace{0.75cm}\tilde{a}_{\text{even}} =\frac{2\beta}{\ell^2}\frac{(\sqrt{\pi}\hspace{.2mm})^{D-1}}{\Gamma(\frac{D-1}{2}) }(r_h^2-\ell^2)r_h^{D-3},\cr \nonumber\\
&&\hspace{0.75cm}\tilde{a}_{\text{odd}} =\tilde{a}_{\text{even}}+2\pi\beta \frac{(D-2)}{(D-3)^2} (-\pi \ell^2)^\frac{D-3}{2},
\eea
where the index even and odd refers to the bulk dimension. The Hawking temperature is then given by
\bea
T_{H}=\frac{1}{\beta}=\frac{1}{4 \pi}\partial_{r} F\big{|}_{r=r_{h}}=\frac{1}{4\pi}\bigg((D-3)\frac{1}{r_h}+\frac{(D-1)}{\ell^2}r_h\bigg),
\eea
which can be used to explain $r_{h}$ in terms of $\beta$ as 
\bea\label{rhD}
r_h=\frac{2\pi\ell^2}{(D-1)\beta}\bigg(1\pm\sqrt{1-\frac{(D-1) (D-3)\beta^2}{4 \pi^{2}\ell^2}}\hspace{1mm}\bigg).
\eea
We continue with the minus sign since it leads to a smaller free energy. Considering the relations among partition function, renormalized on-shell action and energy, i.e.
\bea
\log Z=I_{\text{ren}},\hspace{.5cm} E=-\frac{\partial}{\partial \beta}\log Z, 
\eea 
the energy of the black hole solution (\ref{AdSSchD}) is derived as
\bea\label{ENERGYD}
&& E_{\text{BH}}=\frac{\mathcal{E}}{16\ell^2G}\big(1-\frac{\gamma}{2}+\kappa\hspace{.5mm}\gamma^{\frac{D}{2}-1}\big),\cr \nonumber\\
&&\mathcal{E}_{\text{even}}=(D-1)(D-2)\frac{ (\sqrt{\pi}\hspace{.5mm})^{D-3}}{\Gamma(\frac{D+1}{2})}\hspace{.1mm}(r_{h}^2+\ell^{2})\hspace{.5mm}r_{h}^{D-3},\cr \nonumber\\
&& \mathcal{E}_{\text{odd}}=\mathcal{E}_{\text{even}}-2\ell^2 \frac{(D-2)}{(D-3)^2}(-\pi\ell^2)^{\frac{D-3}{2}},
\eea
where again the index even and odd refers dimension of the bulk spacetime. Furthermore, according to definition of entropy, $ S=\beta E+\log Z$, we have
\bea\label{entropyD}
S_{\text{BH}}=\frac{\mathcal{S}}{G}\big(1-\frac{\gamma}{2} + \kappa\gamma^{\frac{D}{2}-1}\big)\hspace{.5cm}\text{with}\hspace{.5cm}\mathcal{S}=\frac{(D-1)}{4 }\frac{(\sqrt{\pi}\hspace{.3mm})^{D-1} }{\Gamma(\frac{D+1}{2})}r_h^{D-2}.
\eea
In section.\ref{DC}, we showed that absence of Boulware-Deser scalar ghost implies $\gamma=\frac{1}{m^2 \ell^2}(D-2)$. For this value of $\gamma$, the expressions for energy and entropy of black hole solutions (\ref{AdSSchD}) become
\bea
&& E=\frac{\mathcal{E}}{16\ell^2G}\bigg(1-\frac{1}{2m^2\ell^2}(D-2)+\kappa \big(\frac{D-2}{m^2\ell^2}\big)^{\frac{D}{2}-1}\bigg),\cr \nonumber\\
&& S=\frac{\mathcal{S}}{G}\bigg(1-\frac{1}{2 m^2 \ell^2}(D-2)+ \kappa \big(\frac{D-2}{m^2 \ell^2}\big)^{\frac{D}{2}-1}\bigg),
\eea
where  $\mathcal{E}$ and $\mathcal{S}$ are given in Eqs.(\ref{ENERGYD}) and (\ref{entropyD}). Interestingly, these values could be written in terms of  the coefficient of the kinetic term of massless graviton (\ref{AAD}) as 
\bea
E=\frac{\mathcal{E}}{16 G \ell^2}\hspace{.5mm}\mathbb{A}_{0},\hspace{.5cm}
S=\frac{\mathcal{S}}{G}\hspace{.5mm}\mathbb{A}_{0}.
\eea
According to Eqs.(\ref{ENERGYD}) and (\ref{entropyD}), $\mathcal{E}$ and $\mathcal{S}$ are positive in even as well as odd bulk dimensions. Fortunately, the absence of Boulware-Deser scalar ghost and positivity of kinetic term of massless graviton, imply that the energy and entropy of Schwarzschild-AdS black hole solution (\ref{AdSSchD}) are also positive. These positive values for black hole solutions might be the sign that the theory (\ref{bnmgD}) at full non-linear level is also healthy.
\section{Conclusion and Discussion}
The key word in studying field theories is "consistency". The most important criteria which determine the consistency of a  gravitational theory are: 1) Absence of Boulware-Deser ghost, 2) Absence of kinetic ghost, 3) Absence of superluminal modes (i.e.  tachyons)  and 4) Predictability (i.e. absence of  local closed time-like curves).

In this paper we showed that the problem of ghost in critical gravity and its higher dimensional extensions can be resolved by giving dynamics to the symmetric rank two auxiliary field appearing in the action of these theories. The new models, at the linear level around the AdS vacuum, are free of Boulware-Deser ghost, kinetic ghost and tachyonic instability within the particular ranges of parameters. Note that for Lorentz invariant theories the conditions 3 and 4 are equivalent.
Moreover, we showed that the energy and entropy of the AdS-Schwarzschild black hole solutions in our model are positive in the same range of parameters. This might be the sign that the model is free of ghost instabilities at the non-linear level as well. 

A natural and very important question which can be asked is as follows. Is it possible that the Boulware-Deser ghost appears again or the non-tachyonic mode changes to tachyonic one at the full non-linear level? In general, the answer may be Yes, however, it needs to be checked explicitly to assure about the answer No.

Let us remind that the dRGT model and HR bigravity model despite passing the consistency conditions 1 and 2  suffer from violating the conditions 3 and 4 at the full non-linear level. These inconsistencies are shown by the method of characteristics\cite{Deser:2012qx}. In this approach, the absence of superluminal propagating modes means that the characteristic matrix determinant does not vanish anywhere.
Also absence of zero and negative norm states implies that this determinant should be non-degenerate \cite{Johnson:1960vt}. Moreover, the predictability (absence of local CTC) implies that  this determinant shows the lack of space-like characteristic surfaces. All these mean that the characteristic matrix determinant also should be calculated for the theory (\ref{bnmgD}); it is the subject of our future works.

In a different point of view, to explain the cosmological constant problem, the mass of the spin-2 particle in massive theories should be small. The interesting feature of all these models is that this small mass, if able to generate a late-time cosmic acceleration, would be protected against large quantum corrections because of restoring the diffeomorphism symmetry in the small mass limit, $M^{2}\rightarrow 0$.  It is shown that flat and closed Friedmann-Lema$\hat{\text{i}}$tre-Robertson-Walker (FLRW) cosmological solutions do not exist in the dRGT model with a flat reference metric\cite{D'Amico:2011jj}.  Also, its open FLRW solutions (and cosmological solutions with general reference metrics) suffer from either Higuchi \cite{Higuchi:1986py} ghost at the level of linear perturbations or from a new non-linear ghost\cite{DeFelice:2013bxa}. Unlike the dRGT model, its bimetric generalization is able to provide the accelerating solutions\cite{Akrami:2012vf} but unfortunately they also suffer from ghost and/or gradient instabilities\cite{Konnig:2014xva}. Hence, exploring the cosmological solutions for our model (\ref{bnmgD}) is important and can be another subject for our future works.

The crucial point is that the main reason for vanishing of the characteristic matrix determinant, existence of the spacelike characteristic surfaces and absence of the cosmological solutions in dRGT and HR bigravity models is the constraint which removes the Boulware-Deser ghost at the full non-linear level. It is argued that the only possible way to remove the problems caused by this condition is the existence of PM modes \cite{Deser:2012qx} and \cite{Deser:2013eua}. Actually the PM action has an enhanced symmetry which can protect the mass of spin-2 particle against receiving the large non-linear corrections. Unfortunately, this hope is also excluded for  dRGT and HR bigravity theories precisely at the non-linear level\cite{Deser:2013gpa}\footnote{The criticisms of these works can be found in \cite{Hassan:2014vja}.} (even though their linearization has PM mode\cite{Hassan:2013pca}). Interestingly, the theory (\ref{bnmgD}) has PM modes in its spectrum, at least at the linear level. Therefore another important study which should be done on the theory (\ref{bnmgD}) is checking the existence of  PM modes at the full non-linear level that is also the subject of our future works.

Another study which can be done on the theory (\ref{bnmgD}) is understanding the mechanism by which the mass appears in this theory. 
\section{Acknowledgment}
We would like to thank M. Alishahiha for collaboration in the early stage of this paper. We are also grateful to him for useful discussions, comments  and encouragements. We would like to thank S. F. Hassan, A. F. Astaneh, R. Fareghbal, A. Mollabashi, M. R. Mohammdi Mozaffar, F. Omidi, S. F. Taghavi and M.R. Tanhayi for useful discussions. This work is supported by Iran National Science Foundation (INSF). 
\section*{Appendix}
\appendix
\section{Details of Linearization}\label{Details of Linearization}
In this appendix we present the detailed calculations of linearization of
different terms in the action (\ref{bnmgD}). To do this we will consider the following 
perturbations around an arbitrary background in $D$ dimensions 
\bea\label{down}
g_{\mu\nu}=\bar{g}_{\mu\nu}+h_{\mu\nu},\hspace{1cm}
f_{\mu\nu}=\gamma(\bar{g}_{\mu\nu}+\rho_{\mu\nu}),
\eea
where $ \bar{g}_{\mu\nu} $ is the background metric.
Assuming $g^{\mu\nu}$ and $f^{\mu\nu}$ as the inverse tensors corresponding to the metrics $g_{\mu\nu}$ and $f_{\mu\nu}$ respectively, Eq.(\ref{down}) reads to second order as
\bea\label{up}
g^{\mu\nu}=\bar{g}^{\mu\nu}-h^{\mu\nu}+
h^{\mu}_{\lambda}h^{\lambda\nu},\hspace{1cm}f^{\mu\nu}=\gamma^{-1} 
(\bar{g}^{\mu\nu}-\rho^{\mu\nu}+
\rho^{\mu}_{\lambda}\rho^{\lambda\nu}),
\eea
where  $h^{\mu\nu}\equiv \bar{g}^{\mu\alpha}\bar{g}^{\nu\beta}h_{\alpha\beta}$ and  $\rho^{\mu\nu}\equiv \bar{g}^{\mu\alpha}\bar{g}^{\nu\beta}\rho_{\alpha\beta}$. In what follows, we present some formula valid for any arbitrary background

\begin{equation*}
\hspace{-1cm}\sqrt{-g}^{(1)}=\sqrt{-\bar{g}}\hspace{1mm}\dfrac{h}{2},
\end{equation*}
\bea
\sqrt{-g}^{(2)}=\sqrt{-\bar{g}}\hspace{1mm}\frac{1}{8}(h^2-2 h_{\mu\nu}h^{\mu\nu}).
\eea
\bea
&&\Gamma[g]{^{(1)\alpha}}_{\beta\gamma}=\frac{1}{2}\bar{g}^{\alpha\sigma}(h_{\sigma\gamma;\beta}+ h_{\beta\sigma;\gamma}-h_{\beta\gamma;\sigma}),\nonumber\\
&&\Gamma[g]{^{(2)\alpha}}_{\beta\gamma}=-\frac{1}{2}h^{\alpha\sigma}(h_{\sigma\gamma;\beta}+ h_{\beta\sigma;\gamma}- h_{\beta\gamma;\sigma}).
\eea

\begin{equation*}
R[g]^{(1)}_{\mu\nu}=\frac{1}{2} (-{h_{\mu\nu;\alpha}}^{\alpha}-h_{;\nu\mu}+{h_{\mu\sigma;\nu}}^{\sigma}+{h_{\nu\sigma;\mu}}^{\sigma}),
\end{equation*}
\bea
&&R[g]^{(2)}_{\mu\nu}=-\frac{1}{2}{h^{\rho\sigma}}_{;\rho}(h_{\nu\sigma;\mu}+h_{\mu\sigma;\nu}-h_{\mu\nu;\sigma})-\frac{1}{2}h^{\rho\sigma}(h_{\nu\sigma;\mu\rho}+h_{\mu\sigma;\nu\rho}-h_{\mu\nu;\sigma\rho}-h_{\rho\sigma;\mu\nu})\nonumber\\
&&\hspace{1.3cm}+\frac{1}{4}h^{;\alpha}(h_{\nu\alpha;\mu}+h_{\mu\alpha;\nu}-h_{\mu\nu;\alpha})-\frac{1}{2}(h_{\mu\alpha;\rho}{h_{\nu}}^{\rho;\alpha}-h_{\mu\alpha;\rho}{h_{\nu}}^{\alpha;\rho}-\frac{1}{2} {h^{\alpha\rho}}_{;\nu}h_{\alpha\rho;\mu}).
\nonumber\\
\eea
\begin{equation*}
R[g]^{(1)}=-{h_{;\alpha}}^{\alpha}+{h^{\mu\nu}}_{;\nu\mu}-h^{\mu\nu}{R^{(0)}}_{\mu\nu},
\end{equation*}
\begin{eqnarray}
&&R[g]^{(2)}=-{h^{\rho\sigma}}_{;\rho}{h_{\mu\sigma}}^{;\mu}+{h^{\rho\sigma}}_{;\rho}h_{;\sigma}
-h^{\rho\sigma}({h_{\mu\sigma;}}{^{\mu}}_{\rho}+h_{\mu\sigma;}{_{\rho}}^{\mu}-h_{;\rho\sigma}-{h_{\rho\sigma;}}{^{\mu}}_{\mu})\nonumber\\
&&\hspace{1.3cm}-\frac{1}{4}h^{;\alpha}h_{;\alpha}-\frac{1}{2}h_{\mu\alpha;\rho}h^{\mu\rho;\alpha}+\frac{3}{4}h_{\alpha\rho;\mu}h^{\alpha\rho;\mu}+h^{\mu\rho}{h_{\rho}}^{\nu}R^{(0)}_{\mu\nu}.
\end{eqnarray}
\bea
\sqrt{-f}=\gamma^{\frac{D}{2}}\sqrt{-g} (h\rightarrow\rho),\nonumber\\
\Gamma[f]=\Gamma[g](h\rightarrow \rho),\nonumber\\
R[f]_{\mu\nu}=R[g]_{\mu\nu}(h\rightarrow \rho),\nonumber\\
R[f]=\gamma^{-1}R[g](h\rightarrow \rho).
\eea

The zeroth, first and second order of different terms of the action (\ref{bnmgD}) with respect to the above perturbations are given by
\bea\label{firstterm}
&&(\sqrt{-g}R[g])^{(0)}=\sqrt{-\bar{g}}\hspace{1mm}R[g]^{(0)}
,\cr\nonumber\\
&&(\sqrt{-g} R[g])^{(1)}=\sqrt{-\bar{g}}\hspace{1mm}\bigg(-h_{;\mu}
^{~\mu}+h_{\mu\sigma}^{~~;\mu\sigma}-h^{\mu\nu}{R[g]^{(0)}}_{\mu\nu}+\frac{h}{2} R[g]^{(0)} \bigg),\cr\nonumber\\
&&(\sqrt{-g}R[g])^{(2)}=\sqrt{-\bar{g}}\hspace{1mm}
\bigg(+\frac{3}{4}h_{\nu\rho;\mu}h^{\nu\rho;\mu}-
\frac{1}{2}h_{\mu\nu;\rho}h^{\mu\rho;\nu}
+\frac{1}{2}h(h^{\mu\nu}_{~~;\mu\nu}-{h^{;\alpha}}_{\alpha})-\frac{1}{4}h^{;\mu}h_{;\mu}\cr\nonumber\\
&&\hspace{4.1cm}+{h^{\alpha\sigma}}_{;\alpha}(h_{;\sigma}-{h_{\mu\sigma}}^{;\mu})-h^{\rho\sigma}({h_{\mu\sigma;}}^{\mu}_{\rho}+{h_{\sigma\alpha;\rho}}^{\alpha}-h_{;\rho\sigma}-{h_{\rho\sigma;\alpha}}^{\alpha})
\cr\nonumber\\
&&\hspace{4.1cm}+(h^{\mu\rho}{h_{\rho}}^{\nu}-\frac{1}{2} h h^{\mu\nu}){R[g]^{(0)}}_{\mu\nu}+\frac{1}{8}(h^2-2 h_{\mu\nu}h^{\mu\nu})R[g]^{(0)}\bigg),
\cr\nonumber\\
&&\sqrt{-f}R[f]=\gamma^{\frac{D}{2}-1}\sqrt{-g}R[g](h\rightarrow \rho).
\eea
\\

\bea
&&(\sqrt{-g} \tilde{f}^{\mu\nu}G[g]_{\mu\nu})^{(0)}=\dfrac{\gamma(2-D)}{2} \sqrt{-\bar{g}} R^{(0)},\cr\nonumber\\
&&(\sqrt{-g}\tilde{f}^{\mu\nu}G[g]_{\mu\nu})^{(1)}=
\gamma\sqrt{-\bar{g}}\left(\dfrac{2-D}{2} (-h_{;\mu}^{~\mu}
+h_{\mu\nu}^{~~;\mu\nu})+R_{\mu\nu}^{(0)}
(\rho^{\mu\nu}+\dfrac{D-4}{2}h^{\mu\nu})-\dfrac{R^{(0)}}{2}
(\rho+\dfrac{D-4}{2}h)\right),\cr\nonumber\\
&&(\sqrt{-g}\tilde{f}^{\mu\nu}G[g]_{\mu\nu})^{(2)}=
\gamma\sqrt{-\bar{g}}\hspace{.5mm}\bigg(\frac{D-2}{2}\Big[
h^{\nu\sigma}_{~~;\nu} h_{\mu\sigma}^{~~;\mu}-h^{\nu\sigma}_
{~~;\nu}h_{;\sigma}+ 
h_{\nu}^{\sigma} h_{\mu\sigma ;}^{~~~~\mu\nu}
+\frac{1}{4}h^{;\mu}h_{;\mu}\cr\nonumber\\
&&\hspace{3cm}+\frac{1}{2}h_{\mu\sigma;\nu}h^{\mu\nu;\sigma}
-\frac{3}{4}h_{\mu\sigma;\nu}h^{\mu\sigma;\nu}+h^{\nu\sigma}h_{\rho\sigma;\nu}^{~~~~\rho}\Big]
-h^{\nu\sigma}h_{\rho\sigma;\nu}^{~~~~\rho}+
\frac{D-4}{4} h (h_{;\mu}^{~\mu}- h^{\mu\nu}_{~~;\mu\nu})\cr\nonumber\\
&&\hspace{3cm}+\frac{3-D}{2}h^{\mu\nu}(h_{;\mu\nu}+{h_{\mu\nu;\alpha}}^{\alpha})+\frac{D-6}{2} R^{(0)}_{\mu\nu}(\frac{1}{2}h h^{\mu\nu}-h^{\mu\rho}{h_{\rho}}^{\nu})\cr\nonumber\\
&&\hspace{3cm}+\frac{D-6}{8} R^{(0)}(h ^{\mu\nu}h_{\mu\nu}-\frac{1}{2}h^2)+\frac{1}{2}\rho^{\mu\nu}\left(-h_{\mu\nu;\sigma}^{~~~~\sigma}-
h_{;\mu\nu}+2 h_{\mu\sigma;\nu}^{~~~~\sigma}\right)\cr\nonumber\\
&&\hspace{3cm}+\frac{1}{2}\rho(h_{;\mu}^{~\mu}-h^{\mu\nu}_{~~;\mu\nu})+R^{(0)}_{\mu\nu}(\frac{1}{2}\rho h^{\mu\nu}+\frac{1}{2}h\rho^{\mu\nu}-2 \rho^{\mu\rho}{h^{\nu}}_{\rho})+\frac{1}{2}R^{(0)}(h^{\mu\nu}\rho_{\mu\nu}-\frac{1}{2}h\rho)\bigg).\cr\nonumber\\
\eea
\bea
&&(\sqrt{-g} \tilde{f}^{\mu\nu}f_{\mu\nu})^{(0)}=\sqrt{-\bar{g}} D 
\gamma^{2},\:\:\:\:(\sqrt{-g} \tilde{f}^{\mu\nu}f_{\mu\nu})^{(1)}=\sqrt{-\bar{g}} \gamma^{2}
\left(2\rho+\frac{D-4}{2}h\right),\cr\nonumber\\
&&(\sqrt{-g} \tilde{f}^{\mu\nu}f_{\mu\nu})^{(2)}=\sqrt{-\bar{g}} \gamma^{2}
\left(\frac{12-D}{4} h^{\mu\nu}h_{\mu\nu}-4  
h^{\mu\nu}\rho_{\mu\nu}+ \rho^{\mu\nu}\rho_{\mu\nu}+\frac{D-8}{8}h^{2}+h\rho\right).
\nonumber\\
\eea
\bea\label{lastterm}
&&(\sqrt{-g}\tilde{f}^2)^{(0)}=\sqrt{-\bar{g}} D^2 \gamma^{2},\:\:\:\:\:
(\sqrt{-g}\tilde{f}^2)^{(1)}= \sqrt{-\bar{g}}  \gamma^{2}
\left(\frac{D(D-4)}{2}h+2D\rho\right),\cr\nonumber\\
&&(\sqrt{-g}\tilde{f}^2)^{(2)}=\gamma^{2}\sqrt{-\bar{g}}
\left((\frac{D^2}{8}-D+1)h^{2}+(D-2) \rho h+\rho^{2}+\frac{D (8-D)}{4}
h^{\mu\nu}h_{\mu\nu}-2 D h^{\mu\nu}\rho_{\mu\nu}\right).
\nonumber\\
\eea

Putting all of the above expressions in the action and neglecting the boundary terms, the first order terms with respect to the perturbations are as follows
\bea\label{first order equation}
h^{\mu\nu}\Big( R_{\mu\nu}^{(0)}(1-\frac{\gamma}{2}\frac{D-4}{D-2})\Big)+h\Big(\frac{1}{2}R^{(0)}(-1+\frac{\gamma}{2}\frac{D-4}{D-2})+\frac{m^2\gamma^2}{8 }\frac{(D-1) (D-4)}{D-2}+\Lambda_{g}\Big)=0,\cr\nonumber\\
\rho^{\mu\nu}\Big( R_{\mu\nu}^{(0)}(-\frac{\gamma}{D-2}+\kappa \gamma^{\frac{D}{2}-1})\Big)+\rho \Big(\frac{1}{2}R^{(0)}(\frac{\gamma}{D-2}-\kappa \gamma^{\frac{D}{2}-1})+\frac{m^2\gamma^2}{2 }\frac{D-1}{D-2}+\kappa \gamma^{\frac{D}{2}}\Lambda_{f}\Big)=0.
\nonumber\\
\eea
These terms should vanish due to the equations of motion of the background fields. For the remaining of calculations we prefer to consider only $AdS$ and flat backgrounds.  For $AdS$ space in D dimensions, we know that $ R_{\mu\nu}=-\frac{D-1}{l^2}g_{\mu\nu} $ and $ R=-\frac{D(D-1)}{l^2} $. By setting them in the above relation we get the equations of motion in relation (\ref{BeomD}) as below
\bea
&& \Lambda_{g} \ell^2+\frac{1}{2}(D-1)(D-2)-\frac{\gamma}{4}(D-1)(D-4) \Big[1-\frac{m^2}{2 (D-2)}\gamma\ell^{2}\Big]=0,\cr \nonumber\\
&& \Lambda_{f}\gamma\ell^{2}+\frac{1}{2}(D-1)(D-2)+\frac{1}{2\kappa}\frac{(D-1)}{(D-2)}\Big(2-D+\gamma m^2 \ell^{2}\Big)\gamma^{(2-\frac{D}{2})}=0.
\eea
For flat background, the curvature is zero and from (\ref{first order equation}) we have
\bea
\Lambda_{g}+\frac{m^2\gamma^2}{8}\frac{(D-4) (D-1)}{D-2}=0,\cr\nonumber\\
\kappa \gamma^{\frac{D}{2}}\Lambda_{f}+\frac{m^2\gamma^2}{2}\frac{D-1}{D-2}=0.
\eea

We see that the regions corresponding to flat and $ AdS $ solutions are completely disjoint from each other. For example in $D=4$, for the flat case we have $ \Lambda_{g}=0 $ and for the $ AdS $ case we have $  \Lambda_{g}\neq 0 $. So for the special region of parameters where the $ AdS $ solution is valid we cannot have a flat solution and vice versa. Hence, in the region of validity of the $ AdS $ solution we should not be worried about the ghost around the flat solution and vice versa.
 
 For the second order terms, the $ AdS $ case is discussed in the text. In what follows we give the results for the flat background.
By collecting the Eqs.(\ref{firstterm})-(\ref{lastterm}) the perturbed action is similar to the relation (\ref{linearD}) except that the Pauli-Fierz operator on the flat space is defined as 
\bea\label{PFDflat}
&& p^{\mu\nu}(\mathbb{G} q)_{\mu\nu}\equiv -\dfrac{1}{4} p_{\nu\rho ;\mu} q^{\nu\rho ;\mu}+\dfrac{1}{2} p_{\mu\nu ;\rho} q^{\rho\nu ;\mu}-\dfrac{1}{4} p_{;\mu} {q^{\mu\nu}}_{;\nu}-\dfrac{1}{4} q_{;\mu} {p^{\mu\nu}}_{;\nu}
+\frac{1}{4}p_{;\mu}q^{;\mu},
\eea
The dot product defined in the relation (\ref{dot}) is as before with the following parameters 
\bea\label{dotproductflat}
\chi=\frac{-m^2\gamma^2}{4},~~~\xi=\frac{-m^2\gamma^2}{8}\frac{D-3}{D-2}.
\eea

Using the new basis for the fluctuations as (\ref{basis}), the second order action with respect to the perturbations is similar to (\ref{S1D}) with the new definition of Pauli-Fierz operator and dot product as stated above. 
To have the ghost free action we set $ \chi=\xi $ which results to $ D=1 $ and obviously is not acceptable. The only remaining possibility for the ghost-free condition is then $ m^2=0 $ which states that our model around the flat background is physically acceptable just for the massless theory.

However, for $ m^2\neq 0 $ there exists another possibility as $ a_1=a_2=1 $ which leads to $ h=\rho $. This means that two metrics $ g_{\mu\nu} $ and $ f_{\mu\nu} $ are proportional to first order.

Let us continue by setting $ m^2=0 $. To omit the cross term we request $ C=0 $. If we choose $ a_1=1 $ arbitrarily, one gets
\bea
a_2=\frac{-2 \kappa (D-2)\gamma^{\frac{D}{2}-1}+2 \gamma}{2 (D-2)+\gamma (4-D)}.
\eea

In this basis, the theory contains two massless modes provided that the coefficients of the kinetic terms, i.e.
\bea
&&A=\kappa \gamma^{\frac{D}{2}-1}+1-\frac{\gamma}{2},\nonumber\\
&&B=\frac{1}{\Big(2 (D-2)+\gamma (4-D)\Big)^2}\Bigg[ 2 (D-2) \gamma^{D-2} \Big(2 (D-2)-\gamma (D-6)\Big) \kappa^2\nonumber\\
&&+\kappa \bigg(\gamma^{\frac{D}{2}-1} \Big(2 (D-2)+\gamma (4-D)\Big)^2-8\gamma^{\frac{D}{2}+1}\bigg)+2 \gamma^{2}(\gamma-2)\bigg],
\eea
are positive. This implies that
\bea
&&A>0\Rightarrow \kappa >\Big(\frac{\gamma}{2}-1\Big) \gamma^{1-\frac{D}{2}},\nonumber\\
&&B>0\Rightarrow \kappa <\Big(\frac{\gamma}{2}-1\Big)\gamma^{1-\frac{D}{2}} ~~~or~~~ \kappa>\dfrac{2 \gamma ^{3-\frac{D}{2}}}{(2-D) \Big(2 (2-D) +\gamma (D-6)\Big)}.
\nonumber\\
\eea
For example, if we set $ \gamma=1 $ so $ \bar{f}=\bar{g}=\eta_{\mu\nu} $, we conclude that in expansion around flat space the theory would be consistent and has two massless modes with positive kinetic terms if
\bea
m^2=0,~~~ \kappa>\dfrac{2}{D^2-4}. 
\eea
It worth note that, as stated before, the region of parameter space which admits the flat solution is completely disjoint from that of the AdS solution. Hence we do not bother ourselves for finding any overlap region where the ghosts are absent for both solutions. In other words, there exists a definite region where the AdS solution is valid and the perturbations around it are ghost free. On the other hand, there exists another region where the flat solution is valid and does not have ghost perturbations.\textbf{ However, these two regions are completely disjoint from each other.} The theory is safe for AdS as well as flat solution.


\end{document}